\documentclass[letterpaper,12pt]{article}

\usepackage[T1]{fontenc}

\usepackage{geometry}
\usepackage{setspace}
\usepackage{achemso}

\usepackage{graphicx}
\usepackage{float}
\usepackage{bm}
\usepackage{xcolor}
\usepackage[normalem]{ulem}
\newfloat{scheme}{htbp}{los}
\floatname{scheme}{Scheme}
\floatname{chart}{Chart}
\newfloat{graph}{htbp}{loh}

\usepackage{chemformula} 
\usepackage[version = 4]{mhchem} 

\makeatletter
\renewcommand{\@biblabel}[1]{(#1)}
\@ifundefined{bibnumfmt}{}{}
\makeatother
\usepackage{authblk}

\author[1,2]{Zhaohua Tian}
\author[1,3]{Yu Tian$^{*,}$}
\author[1]{Yadi Niu }
\author[1,3]{Qi Liu }
\author[1]{Zihan Mo }
\author[1,6]{Haoyang Zhang }
\author[1,3,4,5,6]{Ying Gu $^{\dagger,}$}

\affil[1]{State Key Laboratory of Artificial Microstructure and Mesoscopic Physics $\&$ Department of Physics, Peking University, Beijing 100871, China\\}
\affil[2]{Department of Physics, Zhejiang Sci-Tech University, Hangzhou 310018, China\\}
\affil[3]{Frontiers Science Center for Nano-optoelectronics $\&$ Collaborative Innovation Center of Quantum Matter $\&$ Beijing Academy of Quantum Information Sciences, Peking University, Beijing 100871, China\\}
\affil[4]{Collaborative Innovation Center of Extreme Optics, Shanxi University, Taiyuan, Shanxi 030006, China\\}
\affil[5]{Peking University Yangtze Delta Institute of Optoelectronics, Nantong 226010, China\\}
\affil[6]{Hefei National Laboratory, Hefei 230088, China}

\title{ Efficient Waveform Capture without Absorption with Synthesis of Complex Frequencies}
\date{*Email: tian-yu@stu.pku.edu.cn\\$^{\dagger}$Email: ygu@pku.edu.cn}

\begin{document}

\maketitle

\begin{abstract}
A optical waveform can be synthesized by complex-frequency waves as well as by real-frequency harmonic waves. While single complex-frequency wave with exponentially rising waveform can be perfectly absorbed in lossless structures. Here, we propose that  diverse optical waveforms can be captured without any absorption through the synthesis of complex frequencies in a lossless system.
The scattering matrix zeros of the system correspond to a set of complex frequencies with exponentially rising waveforms, each of which can be virtually and perfectly absorbed. Thus, an  input waveform, decomposed into these complex frequencies automatically, can be captured without any absorption.
Then, in a well-designed coupled cavity system, various waveforms such as exponentially decaying, Gaussian, rectangular, triangular,  and randomly generated piecewise-linear profiles, are captured with high efficiency. 
The proposed mechanism has potential applications in enhancing light-matter interactions and optical energy storage.

\end{abstract}

\section*{Keywords}
\textit{complex frequency, synthesis of complex frequencies, scattering matrix zeros, light capture, waveform synthesis, virtual perfect absorption}




\section{Introduction}
Light carries both energy and information. Capturing light with  diverse optical waveforms is of great importance for enhancing light-matter interactions, efficient power transfer, signal detection, and photonic qubit storage \cite{PRL2018AluWirelessPowerTrans,Nature2023FlurinDetec,NatCommun2018JulienQubitMemo,RitterNature2012PQST,RMPSolano2019StrongCoupling,NanoLett2021BrooksWGM}. 
In general, when  a waveform is incident in optical structures, the absorption is always accompanied with scattering.
By introducing appropriate loss to modulate the interference of two incident waves, the scattering can be completely eliminated, called coherent perfect absorption \cite{PRL2010StoneCP,ChongNatRewMat2017CPA,AOP2019AnomaliesScatAlu}. In this case, a specific real-frequency wave is perfectly absorbed, whereas this is unachievable for  waves with other frequencies. To overcome the narrowband absorption, white-light cavities \cite{OL2014ScheuerWhiteCav} and exceptional points \cite{PRL2019StoneCPAEP,Science2021LanYangCPAEP,PRL2024StefanEPCPA} are employed for efficient broadband absorption.  
However, the perfect absorption is still limited to single frequency harmonic wave, thus it can not be used to synthesize  diverse input waveforms.
Moreover, these coherent perfect absorption processes are realized in lossy systems, which lead to the complete dissipation of light after capturing, making storage and subsequent use impossible.

\par To overcome the loss, the virtual perfect absorption (VPA) of complex-frequency waves in lossless structures is proposed \cite{Optica2017AluVPA,ACS2020VCC-Alu,Science2025AluComplexExcitation}, where an exponentially rising (ER) waveform corresponding to the scattering matrix zero can be completely absorbed.
Then, by tuning the system to a high-order exceptional point, the exponentially decaying (ED) waveforms are efficiently absorbed \cite{PRA2022StoneEPVPA,PRA2024StoneEPVPA}. 
Besides, real-frequency harmonic waves are absorbed through the dynamic modulation of the imaginary part of scattering matrix zeros \cite{PRAppl2019TretyakovTimeVarying,PRB2020DimitriosDynamicVPA}.
However,  existing VPA is still restricted to a single complex frequency while it can't be used to capture  diverse input waveforms.
Fortunately, from a mathematical perspective,  a broad class of optical waveforms can be synthesized by a set of complex-frequency waves as well as by real-frequency waves \cite{Book2011FrankMMforPhys}.
Accordingly, we propose to synthesize  an input waveform with a series of complex frequencies corresponding to ER waveforms.
If each of these ER waveforms can be virtually absorbed, then the capturing of  input waveforms can be realized.
Since the system is lossless, once the synthesized light is captured, it can be stored and further be utilized for enhancing light-matter interaction and information processing \cite{Kimble2008QuanInter,Science2007Obrien}.

\par In this research, we propose the mechanism of  efficient waveform capture (EWC) without absorption in a lossless system through the synthesis of complex frequencies. 
The scattering-matrix zeros of this system correspond to a series of complex frequencies with ER waveforms, each of which can be virtually and perfectly absorbed. Therefore, an  input waveform, which can be decomposed into these complex frequencies in the lossless system, can be captured without any absorption. 
Furthermore, we have proved the equivalence between capture efficiency and the synthesis fidelity.
Then, using well-designed chain coupled cavities, we demonstrate high-efficient capture of various typical waveforms, including ED, Gaussian, rectangular, triangular, and randomly generated piecewise-linear profiles.
Notably, the  EWC does not require injecting multiple complex frequencies. Instead, the  input waveform is directly decomposed into a set of complex frequencies, each of which corresponds to a VPA process. 
This lossless, passive, and linear capturing approach holds potential applications in enhancing light–matter interaction and nonlinear effects, and efficient signal detection.

\section{Theory}
\begin{figure}[!ht]
 \centering
  \includegraphics*[width=86mm]{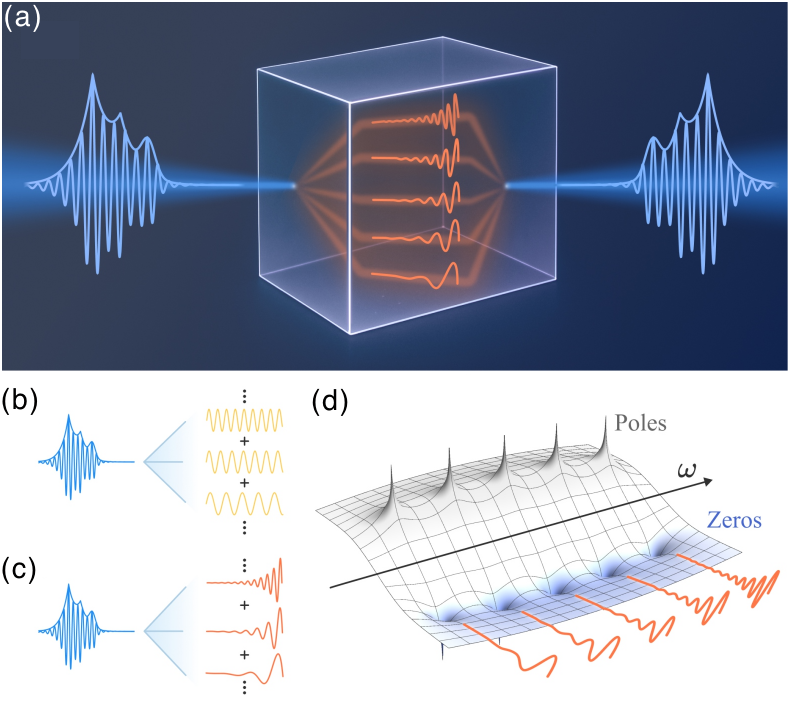}
  \caption{ EWC through synthesis of complex frequencies.
(a) Schematic of  EWC by a lossless system whose scattering matrix zeros correspond to a series of complex frequencies.  An  input waveform can be synthesized from a series of (b)  harmonic real-frequency waves and (c) exponentially rising complex-frequency waves. (d) Eigenvalues of scattering matrix with multiple zeros and poles.}
\label{fig:1}
\end{figure}
The mechanism for  EWC is described as follows [Fig.~\ref{fig:1}]. 
Mathematically, an  input waveform can be decomposed into a series of real-frequency harmonic waves [Fig.~\ref{fig:1}(b)], while it can also be superposed by complex-frequency waves [Fig.~\ref{fig:1}(c)].
Through single scattering matrix zero in a lossless system, the corresponding ER waveform can be virtually perfectly absorbed \cite{Optica2017AluVPA}.
By constructing multiple scattering-matrix zeros corresponding to a series of VPA complex frequencies [Fig.~\ref{fig:1}(d)],  diverse waveforms can be synthesized with these complex frequencies and therefore be captured.
The  EWC is applicable not only to the capture of classical waveforms but also to quantum wavepackets.

\par
\begin{figure}[!t]
\centering
  \includegraphics*[width=86mm]{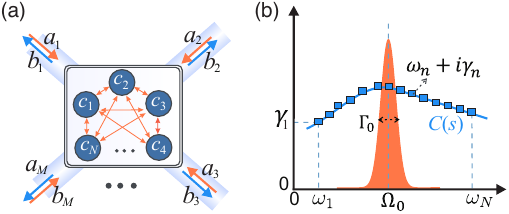}
  \caption{
(a) Schematic of a lossless system with $N$ modes and $M$ channels. (b) The distribution of scattering matrix zeros $\omega_n + i \gamma_{n}$ on the complex plane. The orange region represents the spectral distribution of  the input waveform with the width $\Gamma_0$.}
  \label{fig:2}
\end{figure}

 The system for realizing  EWC can be modeled as a generic structure with $N$ resonant modes and $M$ channels [Fig.~\ref{fig:2}(a)]. The scattering process is described by $|\bm{b}\rangle =\mathbf{S}(\tilde{\omega})|\bm{a}\rangle $, where $|\bm{a}\rangle = [a_1, \ldots, a_M]^T$ and $|\bm{b}\rangle = [b_1, \ldots, b_M]^T$ denote the input and output amplitude vectors, respectively. The scattering matrix is $\mathbf{S}(\tilde{\omega})=\mathbf{I}_{M}-i\mathbf{D}(\tilde{\omega}\mathbf{I}_{N}-\mathbf{H})^{-1}\mathbf{D}^{T}$,
where $\mathbf{H}$ is the system Hamiltonian, $\mathbf{I}$ is the identity matrix, and $\mathbf{D}$ is the real coupling matrix between the resonant modes and the external channels (see Supporting Information).
An eigenvalue of $\mathbf{S}(\tilde{\omega})$ is $\lambda(\tilde{\omega})$ corresponding to the eigenvector $|\bm{d}\rangle$, satisfying $\mathbf{S}(\tilde{\omega})|\bm{d}\rangle =\lambda(\tilde{\omega})|\bm{d}\rangle  $. Perfect absorption occurs when $\lambda(\tilde{\omega})=0$. For the $N$-mode system in Fig. \ref{fig:2}(a), there exist $N$ complex-frequency zeros $\tilde{\omega}_{n}=\omega_{n}+i\gamma_{n} (n=1,2,...,N)$ satisfying $\mathbf{S}(\tilde{\omega}_n)|\bm{d}\rangle =\lambda(\tilde{\omega}_n)|\bm{d}\rangle $ and $\lambda(\tilde{\omega}_{n})=0$. Here, by engineering the coupling matrix $\mathbf{D}$ to be rank one, the cavity modes couple only to one fixed coherent superposition of the external ports. As a result, all these zeros share the same input eigenvector $|\bm{d}\rangle$ (see Supporting Information). For $\tilde{\omega}_n$, its output can be written as
\begin{equation}
  \mathbf{S}(\tilde{\omega}_{n})|\bm{a}\rangle e^{-i\tilde{\omega}_{n} \tau} = \mathbf{S}(\tilde{\omega}_{n})|\bm{d}\rangle e^{-i\tilde{\omega}_{n} \tau}=0
\end{equation}  if $|\bm{a}\rangle=|\bm{d}\rangle$. It is shown that each of the complex frequency waves, featured as an ER waveform, can be virtually and perfectly absorbed. According to the linearity of the scattering response, the output also vanishes for a coherent superposition of these complex-frequency inputs:
\begin{equation}
\begin{aligned}\label{eq:multiCFScat}
	\sum_{n=1}^{N} \alpha_{n} \mathbf{S}(\tilde{\omega}_{n})|\bm{d}\rangle e^{-i\tilde{\omega}_{n}\tau}
	=\sum_{n=1}^{N}\alpha_{n}\lambda(\tilde{\omega}_{n})|\bm{d}\rangle e^{-i\tilde{\omega}_{n}\tau} = 0,
\end{aligned}
\end{equation}
where $\alpha_{n}$ is the coefficient of the $n$-th complex-frequency component. If only one complex-frequency zero exists, it is back to the VPA of single complex frequency \cite{Optica2017AluVPA,ACS2020VCC-Alu,Science2025AluComplexExcitation}.

\par 
We then consider an input waveform superposed by complex frequencies truncated at $t_{0}$
\begin{equation}\label{eq:fCF}
f_{\text{CF}}(\tau,t_{0})=\sum_{n=1}^{N}\alpha_{n} f_{n}(\tau,t_{0}),
\end{equation}
where $f_{n}(\tau,t_0)=\sqrt{2\gamma_{n}}\,e^{-i\tilde{\omega}_{n}(\tau-t_{0})}\Theta(t_{0}-\tau)$ is the normalized complex-frequency basis function defined in $(-\infty,t_0]$. It can be shown through temporal coupled-mode theory \cite{JOSAA2003FanTemporalCMT} that $f_{\text{CF}}(\tau,t_{0})$ can be perfectly captured, with the loading completed at $t_0$ (see Supporting Information). When the complex frequencies are densely distributed around $\Omega_{0}$ and span a range much broader than $\Gamma_{0}$ [Fig. \ref{fig:2}(b)], the summation over these complex-frequency basis functions in Eq. \eqref{eq:fCF} approaches a contour integral (see Supporting Information)
\begin{equation}\label{eq:Laplace}
f_{\text{CF}}(\tau,t_{0})=e^{-i\Omega_{0}(\tau-t_{0})}
\int_{C(s)}\tilde{\alpha}(s)e^{s(\tau-t_{0})} ds\,\Theta(t_{0}-\tau),
\end{equation} 
where $s=-i(\omega-\Omega_{0})+\gamma$ traces the complex-frequency contour $C(s)$. The integral in Eq.~\eqref{eq:Laplace} takes the form of an inverse Laplace representation \cite{Book2011FrankMMforPhys}, through which  a broad class of input waveforms $f_{\text{in}}(\tau)$ can be synthesized from a continuum of complex-frequency components with high fidelity, i.e., $f_{\text{in}}(\tau) = f_{\text{CF}}(\tau,t_0)$. Therefore, an  input waveform with finite bandwidth $\Gamma_{0}$ around the central frequency $\Omega_{0}$ can be  efficient captured by this engineered system.
It should be noted that, although we consider a multiport system, all ports are driven by the same temporal waveform, with their relative amplitudes and phases specified by the scattering-matrix eigenvector $|\bm{d}\rangle$.

In the conventional VPA process, a single scattering zero at a complex frequency is excited by an exponentially rising (ER) waveform \cite{Optica2017AluVPA,ACS2020VCC-Alu,Science2025AluComplexExcitation}, so that the corresponding mode can be perfectly and virtually absorbed. It means that, to realize perfect absorption, inputting an ER waveform is necessary. In practice, the synthesis of ER waveform is challenging \cite{Science2023ZhangSynComplex}, because it will diverge shortly in time. In contrast, our scheme excites a series of complex-frequency zeros simultaneously. Although the input waveform can be understood as a superposition of multiple ER components, the resulting waveform does not contain any divergent part,  which facilitates efficient capture of diverse waveforms.

\par For a practical system with finite modes, an energy-normalized  input waveform ($\int |f_{\text{in}}(\tau)|^2 \allowbreak \mathrm{d}\tau =1$) can be formally decomposed into two orthogonal components  $f_{\text{in}}(\tau)=f_{\text{CF}}(\tau,t_0)+g(\tau)$, where $f_{\text{CF}}(\tau,t_0)$ can be captured, while $g(\tau)$ cannot be, as proven in Supporting Information. Here, we define the synthesis fidelity as the proportion of $f_{\text{CF}}(\tau,t_0)$: $\mathcal{F}=\int \left|f_{\text{CF}}(\tau,t_0)\right|^2\mathrm{d}\tau$. The capturing efficiency $\beta$ is characterized by the maximum energy captured by the system.
Thus the synthesis fidelity of an  input waveform is exactly equal to the capture efficiency, i.e., $\beta=\mathcal{F}$ (see Supporting Information). Therefore,to realize   EWC, the scattering-matrix zeros should be properly engineered to ensure that  input waveforms can be synthesized with high fidelity.

\section{Demonstration of  EWC}
\begin{figure*}[b!]
\centering
  \includegraphics*[width=86mm]{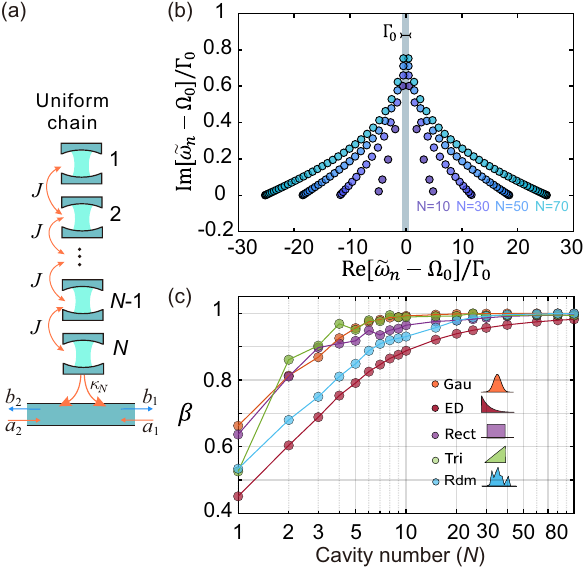}
  \caption{Demonstration of  EWC. (a) Schematic of the periodic cavity chain with uniform coupling strength $J$ between adjacent cavities.  (b) Distribution of the scattering matrix zeros for the cavity number $N=10, 30, 50$ and $70$.
    (c) The capture efficiency as a function of $N$ for five input waveforms. Here $\kappa_{N}=2J$ and $J=0.2\Gamma_{0}N$.}
  \label{fig:3}
\end{figure*}
A chain of coupled cavities is employed to construct the system with multiple scattering matrix zeros [Fig. \ref{fig:3}(a)].
All cavities share the same resonant frequency $\Omega_{0}$ and adjacent cavities are coupled with a uniform strength $J$. The terminal ($N$-th) cavity is coupled to a waveguide, interacting symmetrically with the two propagating modes at a strength of $\sqrt{\kappa_{N}/2}$. The Hamiltonian of the system takes the tridiagonal form
\begin{equation}
\mathbf{H} =
\begin{pmatrix}
\Omega_{0} & J &  &  \\
J & \Omega_{0} & \ddots &  \\
 & \ddots & \ddots & J \\
 &  & J & \Omega_{0}-i\kappa_{N}/2
\end{pmatrix}.
\end{equation}
Here, because only the last cavity is coupled to the external waveguide, the coupling matrix $\mathbf{D} =
\begin{pmatrix}
	0 & 0 & \cdots & \sqrt{\kappa_N / 2} \\
	0 & 0 & \cdots & \sqrt{\kappa_N / 2}
\end{pmatrix}$ has rank one. This ensures that all scattering zeros of the system share the same input eigenvector.
The scattering matrix can be expressed analytically in terms of the system Hamiltonian and the coupling matrix $\mathbf{D}$ (see Supporting Information). Its nontrivial eigenvalue is $\lambda(\omega)=\prod_{n=1}^{N}\frac{\omega-\tilde{\omega}_{n}}{\omega-\tilde{\omega}_{n}^{*}}$ with the corresponding eigenvector $|\bm{d}\rangle=\frac{1}{\sqrt{2}}(1,1)^T$. 
There are $N$ complex frequencies $\tilde{\omega}_{n}$, whose real parts are approximately distributed within $(\Omega_{0}-2J,\Omega_{0}+2J)$, satisfying  $\lambda(\tilde{\omega}_n)=0$. By properly tuning the coupling strength $J$, the distribution of these complex frequencies $\tilde{\omega}_{n}$ can be engineered. 
Fig. \ref{fig:3}(b) shows the corresponding scattering matrix zeros in the complex plane for $N=10,~30, ~50, \text{and } 70$. 
As $N$ increases, these zeros become denser and span a broader spectral region.
This is distinct from the capture mechanism through EP\cite{PRA2022StoneEPVPA,PRA2024StoneEPVPA}, where a few complex-frequency zeros are instead coalesced to achieve efficient capture of a specific ED waveform. 
Leveraging this uniform cavity chain endowed with a series of scattering-matrix zeros, the  EWC can thus be realized.
\par Subsequently, the  EWC is demonstrated by considering five representative input waveforms: ED ($f_{\text{ED}}(\tau)$), rectangular ($f_{\text{Rect}}(\tau)$), triangular ($f_{\text{Tri}}(\tau)$), Gaussian ($f_{\text{Gau}}(\tau)$), and a randomly shaped waveform $f_{\text{Rdm}}(\tau)$ (see Supporting Information). Here, the input waveforms from both channels must match the eigenvector $|\bm{d}\rangle=\frac{1}{\sqrt{2}}(1,1)^T$ of the scattering matrix. The results indicate that the capture efficiencies for all kinds of waveforms approach unity with the increase of $N$ [Fig.~\ref{fig:3}(c)]. A larger $N$ supports a denser and broader distribution of complex frequencies, enabling more accurate synthesis of  input waveforms.
All waveforms can achieve capture efficiencies exceeding 90\% with only 5 cavities and 95\% with 10 cavities, except for the ED waveform.
Even so, when $N=10$ ($N=30$), the capture efficiency of the ED waveform can approach 90\% (95\%). Specially, the capture of ED waveform is of great importance for efficient photonic-qubit detection and storage, as single-photon wavepackets typically possess an ED waveform \cite{MilosNP2016,RenNatCommun2022}. To further clarify the applicability of the proposed mechanism to single-photon wavepackets, a quantum input-output description is provided in the Supporting Information.

\begin{figure*}[t!]
\centering
  \includegraphics*[width=\textwidth]{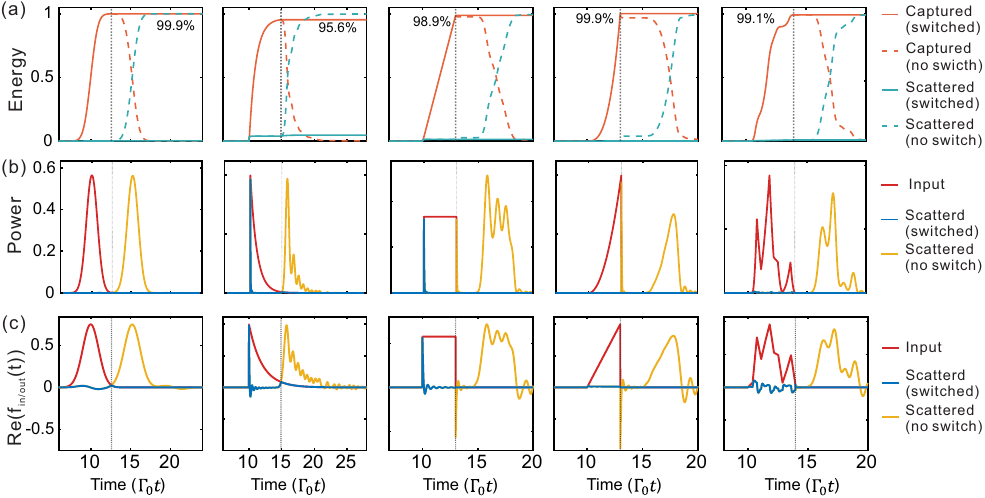}
  \caption{Capture dynamics of five representative input waveforms for the uniform coupled-cavity chain with $N=30$. (a) Energy evolution during the capture process. The orange curves show the intracavity stored energy $E_{\mathrm{cap}}(t)$, and the light-blue curves show the time-integrated scattered energy $E_{\mathrm{scatt}}(t)$. Solid and dashed curves correspond to the cases with and without switching off the coupling between the terminal cavity and the waveguide at the loading completion time $t_{0}$, respectively. (b) Input power, output power after switching off the coupling at $t_{0}$, and output power without switching off the coupling. (c) Real parts of the input waveform and the corresponding output waveforms. The imaginary parts of the output waveforms are zero and are not plotted.}
 \label{fig:4}
\end{figure*}
\par We further examine the capture dynamics for the five representative waveforms at $N=30$ [Fig.~\ref{fig:4}]. The intracavity stored energy is defined as $E_{\mathrm{cap}}(t)=\sum_{n=1}^{N}|c_{n}(t)|^{2}$, while the accumulated scattered energy is $E_{\mathrm{scatt}}(t)=\int_{-\infty}^{t}|f_{\mathrm{out}}(\tau)|^{2}d\tau$. During the loading stage, $E_{\mathrm{cap}}(t)$ increases gradually and reaches its maximum at the loading completion time $t_{0}$, whereas $E_{\mathrm{scatt}}(t)$ remains very small [Fig.~\ref{fig:4}(a)]. This energy balance confirms that nearly all incident energy is transferred into the coupled-cavity chain rather than being scattered away.
After $t_{0}$, if the system remains passive, the captured energy is re-emitted back into the waveguide through emission (dashed lines). The transient confinement of the captured energy is nevertheless useful for applications such as enhanced nonlinear interactions and photodetection. If, instead, the captured energy needs to be stored, the system requires an active external switch that decouples the terminal cavity from the waveguide, i.e., $\kappa_{N} = 0$ (solid lines). Figure~\ref{fig:4}(b) shows the input power $P_{\mathrm{in}}(t)=|f_{\mathrm{in}}(t)|^{2}$ and output power $P_{\mathrm{out}}(t)=|f_{\mathrm{out}}(t)|^{2}$, where the output power is strongly suppressed during the loading interval. Figure~\ref{fig:4}(c) further compares the real parts of the input and output waveforms. For waveforms with sharp temporal features, such as the rectangular and ED waveforms, a small output peak appears at the initial rising edge. However, this peak exists only over a very short time interval and rapidly disappears, while the subsequent loading process remains nearly scattering-free.

\begin{figure}[!t]
\centering
  \includegraphics*[width=86mm]{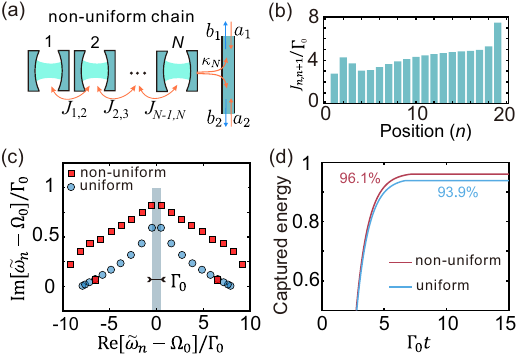}
  \caption{Capture of ED waveform. (a) Schematic of non-uniform cavity chain.  (b) The coupling strength $J_{n,n+1}$ between adjacent cavities with $\kappa_{N}=20\Gamma_{0}$. (c) The distribution of scattering matrix zeros for a uniform chain (blue circles) and a non-uniform chain (red squares). (d) Capturing process of an ED waveform in a uniform chain (blue curves) and non-uniform chain (red curves).}
 \label{fig:5}
\end{figure}

\par To reduce the required number of cavities, a non-uniform coupled-cavity chain can be employed to precisely tailor the distribution of scattering matrix zeros [Fig.~\ref{fig:5}(a)]. We take the ED waveform capture as an example. With the coupling strengths $J_{n,n+1}$ between adjacent cavities [Fig.~\ref{fig:5}(b)], the resulting distribution of scattering-matrix zeros exhibits a clear deviation from the uniform chain case [Fig.~\ref{fig:5}(c)].
With only 20 cavities, the capturing efficiency can exceed 96.1\%, surpassing the uniform chain case of 93.9\% [Fig.~\ref{fig:5}(d)].  While for realizing the same capture efficiency, the uniform chain needs 30 cavities. 
Therefore, with engineering the coupling strengths of non-uniform chain of cavities, higher capture efficiencies can be realized with less cavities.

\section{Discussion}
We first discuss the  applicability conditions  of the capture mechanism. The capture efficiencies in Fig.~\ref{fig:3}(c) are obtained for a fixed input bandwidth $\Gamma_{0}$ and a coupling strength $J$ that scales linearly with the cavity number. To examine the applicability conditions of  EWC, we vary these two parameters for different $N$ in the Supporting Information. The results show that, for each fixed $N$, there exist optimal values of $\Gamma_{0}$ and $J$ corresponding to a balance between sufficiently broad spectral coverage and sufficiently dense zero spacing of the scattering-matrix zeros. As the number of cavities increases, high-efficiency capture can be achieved over a broader parameter region. We also use chirped Gaussian waveforms as an example to examine the influence of temporal duration at fixed spectral bandwidth. The results show that chirp-induced temporal broadening can reduce the capture efficiency, while this influence gradually becomes weaker as the number of cavities $N$ increases.

We then discuss the influence of  parameter deviations and loss of cavities on the  EWC. 
When the coupling strengths of adjacent cavities deviate from the designed, there will be a shift of the distribution of scattering matrix zeros. This will reduce the synthesis fidelity of complex frequencies,  leading to the decrease of capture efficiency. To quantify this effect, we take the uniform chain cavities in Fig. \ref{fig:3}  with $N=30$ as an example. When the coupling strength $J$ has a variation of 10\%, the average capture efficiency drops by less than 1\%, exhibiting strong robustness against the coupling strength deviations (see Supporting Information). Additionally, the capture efficiency is also immune to cavity detuning to some extent. For example, when there is a detuning $2\Gamma_{0}$ between the cavity and the central frequency of input waveforms, the capture efficiency will reduce by less than 1\% (see Supporting Information). Another cause to reduce the capture efficiency is cavity loss, which will cause part of the energy to dissipate during the capturing process. To quantitatively assess this impact, we use the uniform chain structure in Fig. \ref{fig:3} as an example. For the loss $\Gamma_{\text{C}}$ across all cavities, if $\Gamma_{\text{C}}/\Gamma_{0}<0.01$, the average capture efficiency keeps at 96\%, while for $\Gamma_{\text{C}}/\Gamma_{0}<0.05$, it can still maintain an efficiency above 90\% (see Supporting Information). Moreover, loss eventually leads to the dissipation of captured energy, making long-term storage difficult.

 If the captured energy is to be stored, the precise and fast switching of the coupling between the cavity and waveguide is required. To quantify the influence of non-ideal switch, we model the dynamic switch on/off process using a logistic function. We calculate the influence of finite switching duration and timing jitter on the capture efficiency, retrieval efficiency, and modulation-induced sidebands. We have also estimated the influence of switching-related loss on the total retrieval efficiency by introducing an additional efficiency factor for each switching operation (see Supporting Information).   The results show that the influence of non-ideal switching is negligible as long as the switching duration and timing jitter are kept below  $0.1/\Gamma_{0}$. It is worth emphasizing that this dynamic switching plays a fundamentally different role from that in previously proposed time-varying capture schemes \cite{PRAppl2019TretyakovTimeVarying,PRB2020DimitriosDynamicVPA}. In those schemes, the time-dependent modulation is the key mechanism that enables waveform capture and therefore requires a precisely designed temporal profile throughout the absorption process. In contrast, waveform capture in our scheme is entirely passive. Dynamic modulation is only applied after the capture is completed to isolate the stored energy, and its temporal profile is not critical as long as the switching is sufficiently fast.

\par  Finally, we discuss realistic parameters for implementing  EWC in chain-coupled cavity systems. For microwave signals with a central frequency $\Omega_{0}/2\pi\sim 10~\mathrm{GHz}$ and a spectral width $\Gamma_{0}/2\pi\sim 1~\mathrm{MHz}$ \cite{PRXQuan2025NakamuraMicrowave}, the required inter-cavity coupling strength, $J/2\pi$, lies in the range of $10$–$100~\mathrm{MHz}$, which is readily achievable in circuit-QED experiments \cite{RMP2021AndreasCircuitQED}.  Quality factors of $10^{6}\sim 10^{7}$ have been demonstrated for microwave cavities \cite{APL2015MicrowaveHighQ,PRB2016MicrowaveHighQ}, allowing the condition $\Gamma_{C}/\Gamma_{0}<0.01$ to be readily satisfied. These parameters indicate that the capture of  microwave pulses at hundreds  of nanosecond-scale is experimentally feasible.
For optical pulses, taking $\Omega_{0}/2\pi\approx385~\mathrm{THz}$ and $\Gamma_{0}/2\pi\approx160~\mathrm{MHz}$ as representative values \cite{MilosNP2016,RenNatCommun2022,NanoLett2017KimHybrid}, the required inter-cavity coupling strength $J/2\pi$ ranges from approximately $10~\mathrm{MHz}$ to $1~\mathrm{GHz}$. Such cavity chains can be implemented using microring resonators on integrated photonic platforms \cite{LipsonOptica2017,JohnNP2021SiN,NatCommun2021OpticalHighQ}. Coupling strengths between adjacent resonators can reach tens of gigahertz and can be conveniently engineered by adjusting the inter-resonator gap \cite{BergmanJLT2018MRRDesign,TianPRAppQST}. The more demanding requirement in the optical regime is the cavity quality factor. Maintaining an average capture efficiency above 96\% requires Q $\sim 10^{8}$, which, although demonstrated in state-of-the-art experiments \cite{JohnNP2021SiN, NatCommun2021OpticalHighQ}, remains at the frontier of current fabrication capabilities. For the more routinely achievable Q $\sim 10^{7}$ \cite{LipsonOptica2017}, an average capture efficiency above 90\% can still be obtained. For the on-chip integration of large-scale optical cavity chains, fabrication-induced disorder needs to be carefully controlled. Resonance-frequency shifts and coupling deviations can be compensated using post-fabrication or in situ tuning techniques, including photoinduced resonance tuning, electro-optic tuning, trimming, and MEMS actuation \cite{OE2015LibNo3Tuning,JelenaAPL2008,WuOE2006,GleasonOL2005}. It should be noted that the representative experimental values discussed above correspond to separate demonstrations of individual parameters. Integrating these ingredients simultaneously into a switchable, low-loss, and well-controlled multi-cavity chain remains experimentally challenging, especially in optical platforms. More detailed parameter requirements and estimates across different experimental platforms are provided in the Supporting Information.

\section{Conclusion}
We have proposed that  diverse optical waveforms in the lossless system can be captured  without absorption through the synthesis of complex frequencies.  Here, the waveform to be captured is assumed to be an finite-energy pulse whose energy is predominantly confined within finite temporal and spectral ranges. Though the input waveform is superposed with a set of ER waveforms, but it does not contain any diverging part, so it facilitates the  efficient capture of a borad class of waveforms.
By constructing the system with multiple scattering-matrix zeros corresponding to perfectly-absorbed complex frequencies,  the input waveforms can be synthesized by these complex frequency waves, therefore can be captured. 
We have demonstrated the capture of various waveforms using coupled-cavity chain system, where the capture efficiencies can approach to unity as the increase of cavity number. 
The proposed  EWC mechanism has potential applications in enhancing light–matter interactions, efficient optical energy storage,  and signal capture and detection.

\section*{Acknowledgements}
This work is supported by the the financial support from the National Natural Science Foundation of China under Grant No. 12474370 and No. U25D9003 and the Quantum Science and Technology-National Science and Technology Major Project No. 2021ZD0301500.

\section*{Supporting information}
The Supporting Information includes: Theoretical framework for waveform capture; Perfect capture of superpositions of complex frequencies;  Integral representation of superpositions of complex frequencies; Equivalence between capture efficiency and synthesis fidelity; Analytical scattering matrix zeros of a coupled-cavity chain; Analytical scattering-matrix zeros of a two-mode system; Capture of five representative waveforms using a coupled-cavity chain; Quantum input–output description of single-photon wavepacket capture; Applicability conditions for the  EWC mechanism; Influence of the parameter deviations; Influence of the cavity loss; Influence of non-ideal switching on/off of the coupling; Realistic platform-level feasibility estimate of  EWC.


\end{document}